# A Caveat for Single-Molecule Magnetism: Non-linear Arrhenius Plots

Christos Lampropoulos, Stephen O. Hill*, and George Christou*

Single-molecule magnets (SMMs) are molecules that below a certain temperature ($T_B$) function as individual nanoscale magnetic particles, exhibiting magnetization hysteresis loops.[1] As such, they represent an alternative and molecular (bottom-up) route to nanomagnetism, complementing the top-down approach to traditional magnetic nanomaterials.[2] SMMs also exhibit fascinating quantum behavior such as quantum tunneling of the magnetization (QTM)[3] and quantum phase interference (QPI),[4] showing that they are truly mesoscale entities straddling the classical/quantum divide. The barrier causing slow magnetization relaxation arises from a combination of a large ground state spin ($S$) and easy-axis anisotropy (negative zero-field splitting parameter, $D$). The most studied SMMs are the [$Mn_{12}O_{12}(O_2CR)_{16}(H_2O)_4$] family with $S$ = 10 ground states, and their derivatives,[5] while in the recent years others have been discovered.[6] Ac magnetic susceptibility studies are a convenient method of assessing whether a compound might be an SMM; frequency-dependent out-of-phase ($\chi''_M$) signals are indicative of the superparamagnet-like properties of an SMM. The variation in signal position with ac frequency can then be used as a source of rate vs $T$ kinetic data, because the $\chi''_M$ peak maximum is the temperature at which the angular frequency of the oscillating field equals the rate (1/$\tau$, where $\tau$ is the relaxation lifetime) of spin vector reversal. This allows construction of a ln(1/$\tau$) vs 1/$T$ plot based on the Arrhenius relationship (eq 1), the behavior expected of a thermally-activated process over a single-barrier.

$$(1/\tau) = (1/\tau_0)\exp(-U_{eff}/kT) \qquad (1)$$

[a] Dr. C. Lampropoulos, Prof. Dr. G. Christou
Department of Chemistry
University of Florida
Gainesville FL 32611-7200 (USA)
Fax: (+1)352-392-8757
E-mail: christou@chem.ufl.edu

[b] Prof. Dr. S. O. Hill
Department of Physics and National High Magnetic Field Laboratory
Florida State University
Tallahassee, FL 32310 (USA)

Supporting information for this article is available on the WWW under http://www.chemphyschem.org or from the author.

From the slope of the straight line can be obtained the effective barrier to relaxation ($U_{eff}$), and from the intercept the pre-exponential term (1/$\tau_0$). Adherence to the Arrhenius relationship has been one defining property of a SMM, reflecting the low-dimensional origin of its magnetic properties rather than 3-D interactions and long-range magnetic order. Deviations from linearity have been noted in the ln(1/$\tau$) vs 1/$T$ plot for [$Mn_{12}O_{12}(O_2CMe)_{16}(H_2O)_4$]·2MeCO$_2$H·4H$_2$O (**1**) when very high ac frequencies (from several kHz up to 1 MHz) were employed.[7,8] In the present work, we show that non-linear Arrhenius behavior is evident even for data collected in the more typical frequency ($\upsilon$) range (5 - 1500 Hz), and that the apparent $U_{eff}$ obtained is thus distinctly dependent on the range of data constrained to a linear fit to the Arrhenius equation, giving erroneously large apparent $U_{eff}$ values. This conclusion will be supported by comparisons of $U_{eff}$ values with those from high-frequency EPR (HFEPR) data ($U_{EPR}$).

The upper limit to the true barrier to magnetization reversal in SMMs can be calculated with knowledge of the spin Hamiltonian parameters, and a reliable method to obtain these is HFEPR spectroscopy. This gives what we shall henceforth refer to as $U_{EPR}$, and, as stated, this represents a theoretical upper limit to the barrier.[9] However, this is not necessarily the same as the effective barrier, $U_{eff}$, measured by kinetic (ac) studies because there may be QTM through higher energy $M_S$ levels of the ground state $S$ multiplet, thereby effectively cutting-off part of the true barrier.[3,9] The slope of an Arrhenius plot has thus been an invaluable way to determine $U_{eff}$ experimentally. We have recently been carrying out determinations of $U_{EPR}$ for various SMMs, and have been encountering significant and puzzling differences between them and the corresponding $U_{eff}$ values obtained from linear Arrhenius plots using ac $\chi''_M$ vs $T$ data obtained with 5 - 1500 Hz frequencies; these studies consistently give $U_{eff}$ > $U_{EPR}$ rather than the other way round.[9] We have therefore re-examined data for many compounds, and we have come to the conclusion that even ln(1/$\tau$) vs 1/$T$ plots constructed using data collected with frequencies <1500 Hz data exhibit noticeable curvature, and that fits to eq 1 are thus inappropriate!

To better probe this, we have collected data at 25 frequencies <1500 Hz for [$Mn_{12}O_{12}(O_2CCH_2Bu^t)_{16}(MeOH)_4$]·MeOH (**2**),[10] and [$Mn_{12}O_{12}(O_2CCH_2Br)_{16}(H_2O)_4$]·4CH$_2$Cl$_2$ (**3**),[11] which (like **1**) crystallize in tetragonal space groups with the Mn$_{12}$ molecules having high (axial, $S_4$) symmetry. The resulting ln(1/$\tau$) vs 1/$T$ plot for **2** is shown in Figure 1, and the usual practice of fitting this to a straight line (eq 1) gives $U_{eff}$ = 72.0(5) K and $\tau_0$ = 4.4 x 10$^{-9}$ s (solid line in Figure 1, top). However, the fit is clearly not great owing to the curvature. If it is assumed the high $T$ (high $\upsilon$) data represent a deviation from otherwise linear behavior,[8] then the low $T$ data (< 600 Hz) can be fit to give $U'_{eff}$ = 69.5(5) K and $\tau_0$ = 7.1 x 10$^{-9}$ s; the high $T$ data (≥ 600 Hz) can themselves also be fit to eq 1 to give $U''_{eff}$ = 81.4(7) K and $\tau_0$ = 1.1 x 10$^{-9}$ s (solid lines in Figure 1, bottom). However, this approach totally ignores the overall curvature, which also makes it impossible to decide objectively where data separation should be made for a two-line fit; this, of course, affects the resulting $U'_{eff}$ calculated from the slope.[12]

For comparison, single-crystal HFEPR spectra on **2** were measured in the 50-360 GHz range, with the field along the easy-axis of the Mn$_{12}$ molecule ($B//z$).[13] Fitting of the resulting data gave $U_{EPR}$ = 67.3(13) K,[14] noticeably smaller than $U_{eff}$ = 72.0 K from the linear fit (Figure 1, top), and closer to the $U'_{eff}$ = 69.5 K of the low $\nu$ (low T) fit. The latter supports the higher $T$ (higher $\nu$) data as being the main source of the deviation from the Arrhenius law and the resulting discrepancy between $U_{EPR}$ and $U_{eff}$. Complex **3** gave similar results: $U_{eff}$ = 77.6(4) K, $U_{EPR}$ = 68.3(13) K, $U'_{eff}$ = 75.6(5) K, $U''_{eff}$ = 83.4(4) K.[12] Therefore, the picture that emerges is that fits of the data to Arrhenius eq 1, as one or two linear regions, are either inappropriate or subjective, or both. Thus, they are unsatisfactory for obtaining reliable $U_{eff}$ values, or ones that can be confidently compared between different workers who will likely use different frequency sets and thus weight differently the low and high frequency data points in a linear fit.

To accommodate the high $T$ (high $\nu$) deviations, the modified Arrhenius relationship of eq 2 was employed, comprising a double-exponential function involving two barriers $U_1$ and $U_2$, and corresponding $\tau_{01}$ and $\tau_{02}$, respectively.

$$(1/\tau) = (1/\tau_{01})\exp(-U_1/kT) + (1/\tau_{02})\exp(-U_2/kT) \quad (2)$$

An excellent fit of the $\ln(1/\tau)$ vs $1/T$ data for **2** to eq 2 was obtained (solid line in Figure 2, top) with $U_1$ = 63(1) K, $\tau_{01}$ = 2.8(6) × $10^{-8}$ s, $U_2$ = 102(4) K, and $\tau_{02}$ = 1.3(7) × $10^{-10}$ s. $U_1$ is reasonable for a $Mn_{12}$ SMM and is close to $U_{EPR}$ (67.3 K); as already stated, it is reasonable for $U_{eff}$ < $U_{EPR}$ due to tunneling below the top of the EPR barrier. In Figure 2 (bottom) is the deviation $\Delta\ln(1/\tau_1)$ of the data from the linear plot resulting from the first term in eq 2 and the $U_1$ and $\tau_{01}$ values; the deviation is non-zero at all $T$ ($\nu$) values, i.e. the second term affects the relaxation kinetics even at the lowest frequencies.

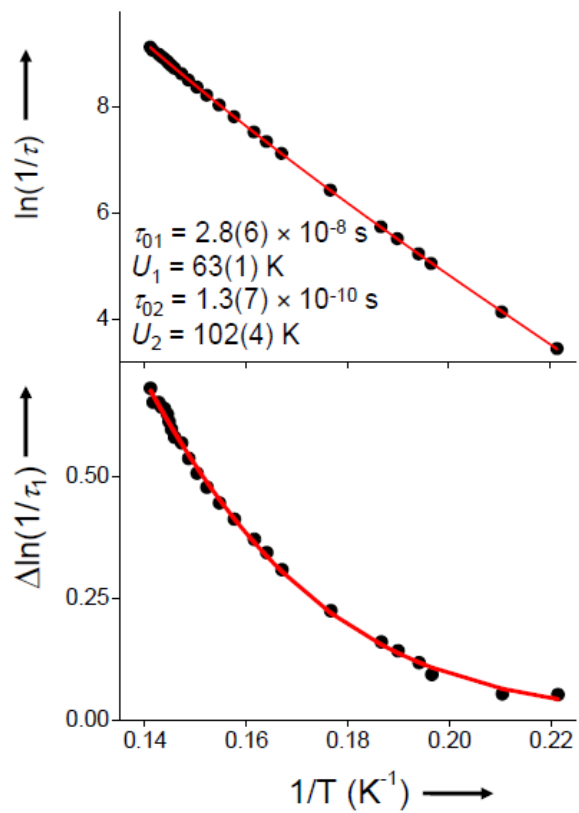

Figure 2. Plot of $\ln(1/\tau)$ vs $1/T$ for **2**. (top) The solid line is the fit to the double-exponential function of eq 2; (bottom) the deviation $\Delta\ln(1/\tau_1)$ of the data from the straight line given by the first term of eq 2.

Eq 2 represents a minimal description of what we believe to be behind the curvature of the $\ln(1/\tau)$ versus $1/T$ plots. It consists of a sum of two relaxation processes: the first term corresponds to relaxation within the $S$ = 10 ground state, and the second term represents relaxation via excited $S$ states. Under this description, $U_1$ is the real kinetic relaxation barrier of the $S$ = 10 $Mn_{12}$ molecule in its ground state, and $U_2$ reflects the 'average' excitation energy to a large group of excited states that provide more efficient relaxation pathways, i.e. a faster pre-exponential factor $1/\tau_{02}$. Relaxation through excited states will involve processes in which spins within the cluster flip independently of each other, i.e. the notion of a rigid spin $S$ = 10 breaks down at higher temperatures. There is clearly only a single relaxation process over the $S$ = 10 barrier for the rigid spin, whereas there will be a huge number of relaxation pathways involving independent spin-flips. We propose that this is the reason for the significantly faster pre-exponential factor.

While the second term in eq 2 is phenomenological, the first does accurately reflect the slower relaxation within the ground state. Consequently, eq 2 correctly captures the low-temperature limiting behavior of the relaxation dynamics, and we thus believe that $U_1$ provides a reliable estimate of $U_{eff}$ that is not dependent on factors such as the choice of ac frequencies. The value of $U_2$, on the other hand, should not be taken too seriously, although it does agree quite well with the spectroscopically determined locations of excited states.[11a] A very large $U_{eff}$ = 168 K (and $\tau_0$ = 1.0 × $10^{-11}$ s) obtained from a fit to eq 1 of $\ln(1/\tau)$ vs $1/T$ data collected in the 100 Hz to 1 MHz frequency range (and $T$ up to 20 K) suggests even faster pathways at higher energies.[7] Thus, an exact description should probably contain stretched exponentials to capture the high $T$ behavior.

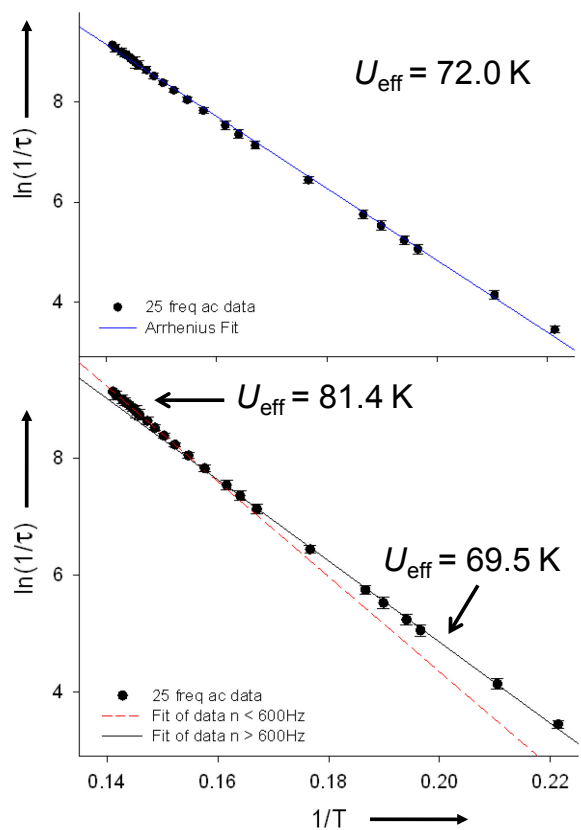

Figure 1. Plot of the natural logarithm of the magnetization relaxation rate vs $1/T$ for **2**: (top) the solid line is the fit of all the data to eq 1; (bottom) lines are the separate (two-line) fits of data at > 600 Hz (dashed red line) and < 600 Hz (solid black line) to eq 1, and are extended to emphasize the different slopes.

In summary, the magnetization relaxation of Mn$_{12}$ SMMs does not follow a simple Arrhenius law, even for data collected at < 1500 Hz, due to contributions to the relaxation rate of pathways via excited $S$ states with a larger thermal barrier but much faster spin reversal rates. We have provided a double-exponential modification of the Arrhenius equation that we consider a superior means to obtain the true barrier $U_1$ of an SMM in its ground state, which also avoids subjective decisions about which data to employ in a linear Arrhenius plot (eq 1) and thus making comparisons between different data sets from different groups and techniques more reliable. Preliminary studies of some other Mn$_x$ SMM families are leading to the same conclusions as for Mn$_{12}$, i.e. non-linear Arrhenius plots at low frequencies are a general property of SMM relaxation kinetics. Note that a modified Arrhenius equation was also needed to fit the spin crossover relaxation rates in light-induced thermal hysteresis studies of certain Fe$^{II}$ compounds, where strong cooperative effects between neighboring molecules lead to self-acceleration effects.[15] Also note that the present results suggest published $U_{eff}$ values of SMMs obtained from Arrhenius plots using ac susceptibility data are consistently overestimated, in some cases by ~10% or more, e.g. the Mn$_6$ family.[16,17] It also appears that accurate comparisons of $U_{eff}$ values that have been obtained from ac data using eq 1 are not possible unless the data were collected in the same frequency ranges. Without doubt, there is a need for greater theoretical understanding in this area.

## Experimental Section

In order to measure the AC magnetic susceptibility of pristine (wet) samples, the crystalline material was first removed from the mother liquor, dried with tissue paper, and rapidly transferred to an analytical balance for accurate weight measurement. Then, within one minute, the crystals were carefully embedded in eicosane within a gelatin capsule in order to ensure retention of the solvent of crystallization. Measurements of the in-phase ($\chi'$) and out-of-phase ($\chi''$) AC susceptibility were made in the 5 - 1500 Hz frequency range using a Quantum Design MPMS-XL magnetometer.

HFEPR measurements were performed on single crystals at various frequencies in the 50 to 400 GHz range using a sensitive cavity perturbation technique and a Millimeter-wave Vector Network Analyzer (MVNA) described elsewhere[13]. The crystals were quickly transferred from the mother liquor and coated with silicone grease in order to avoid solvent loss. The samples were also initially cooled under atmospheric helium gas, with a total transfer time from the mother liquor to the cryostat of just 10-15 minutes. The magnetic field was provided by a 7 T horizontal-bore superconducting magnet associated with a Quantum Design PPMS magnetometer. The horizontal-bore magnet facilitates in-situ rotation of the cavity relative to the applied field. The cavity additionally permits further rotation of the sample about an orthogonal axis (in-situ), thereby enabling collection of data for both easy-axis ($B//c$-axis) and hard-plane ($B \perp c$-axis) orientations. Sample alignment is first achieved by locating extrema among plots of spectra recorded at many different field orientations; once aligned, multi-frequency measurements are performed in order to provide data sets which maximally constrain the ZFS parameters.

## *Acknowledgements*

*This work was supported by NSF (Grants CHE-0414555 and DMR-0804408)*

# SUPPORTING INORMATION

## A Caveat for Single-Molecule Magnetism: Nonlinear Arrhenius Plots

*Christos Lampropoulos, Stephen O. Hill\*, and George Christou\**

**Table S1.** Variation of the calculated $U'_{eff}$ and $U''_{eff}$ for complex **2**, considering different frequency cutoff points for the two linear fits to eq 1.

| Frequency cutoff point (Hz) | $U'_{eff}$ (K) | $U''_{eff}$ (K) |
|---|---|---|
| 1150 | 71.4 | 75.7 |
| 1000 | 70.8 | 79.7 |
| 800  | 70.3 | 81.1 |
| 600  | 69.5 | 81.4 |
| 400  | 69.2 | 80.8 |
| 250  | 68.4 | 78.8 |

Dr. C. Lampropoulos, Prof. Dr. G. Christou
Department of Chemistry
University of Florida
Gainesville FL 32611-7200 (USA)
Fax: (+1)352-392-8757
E-mail: christou@chem.ufl.edu

Prof. Dr. S. O. Hill
Department of Physics and National High
Magnetic Field Laboratory
Florida State University
Tallahassee, FL 32310 (USA)

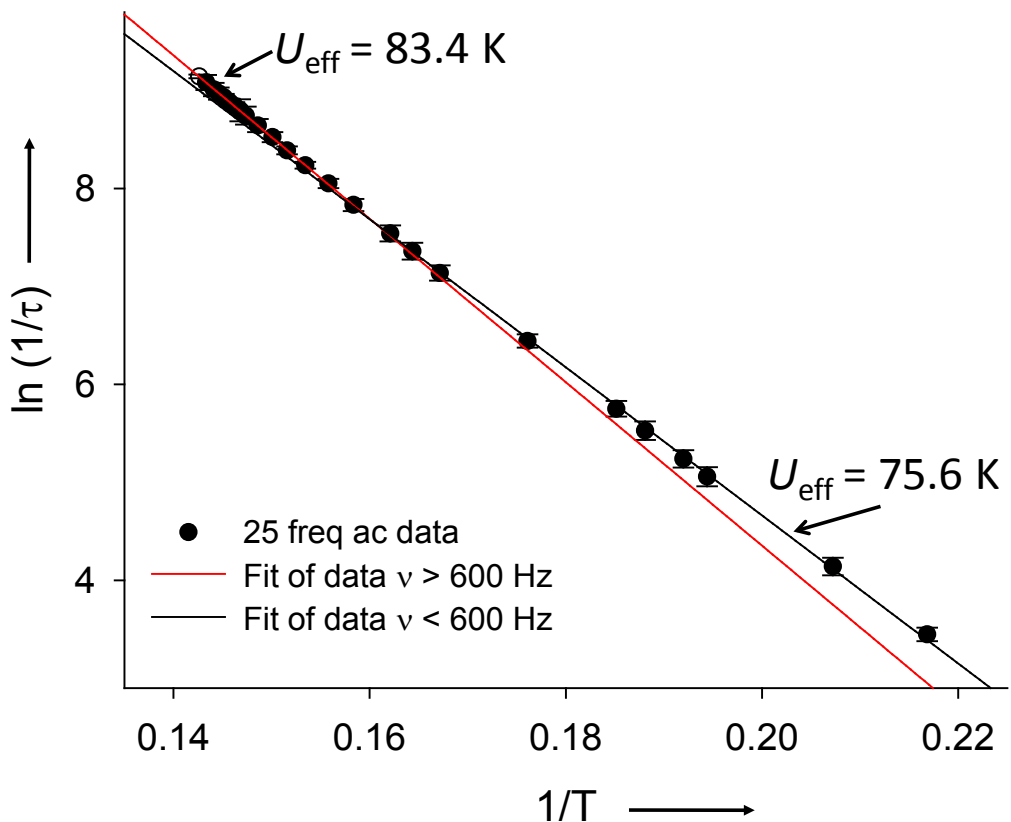

*Figure S1.* Plot of the natural logarithm of the magnetization relaxation rate vs 1/T for **3**: solid lines are the separate (two-line) fits of data at > and < 600 Hz to eq 1; dashed lines are extrapolations of the solid lines.

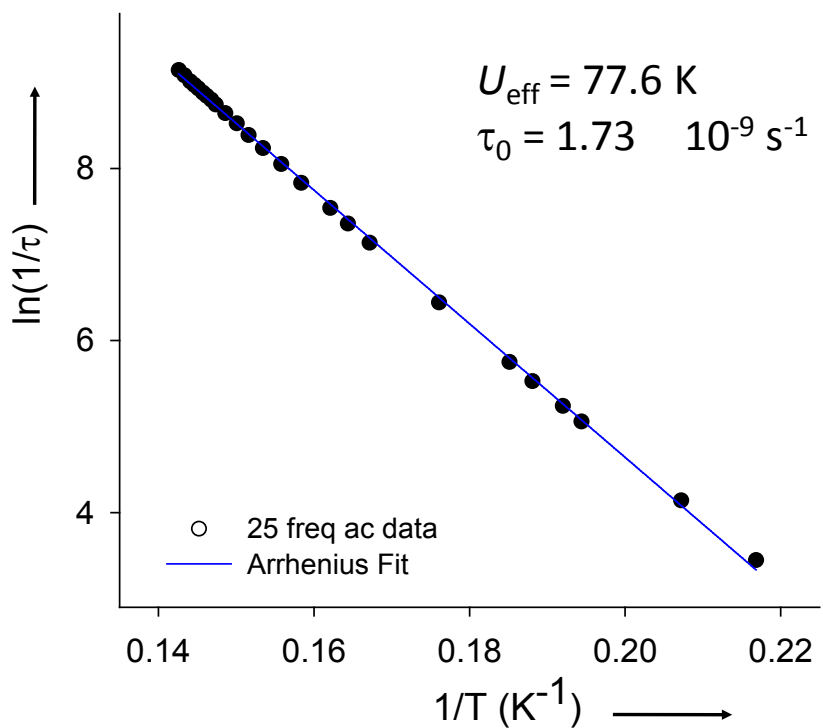

*Figure S2.* The fit of all the data for **3** to eq 1.

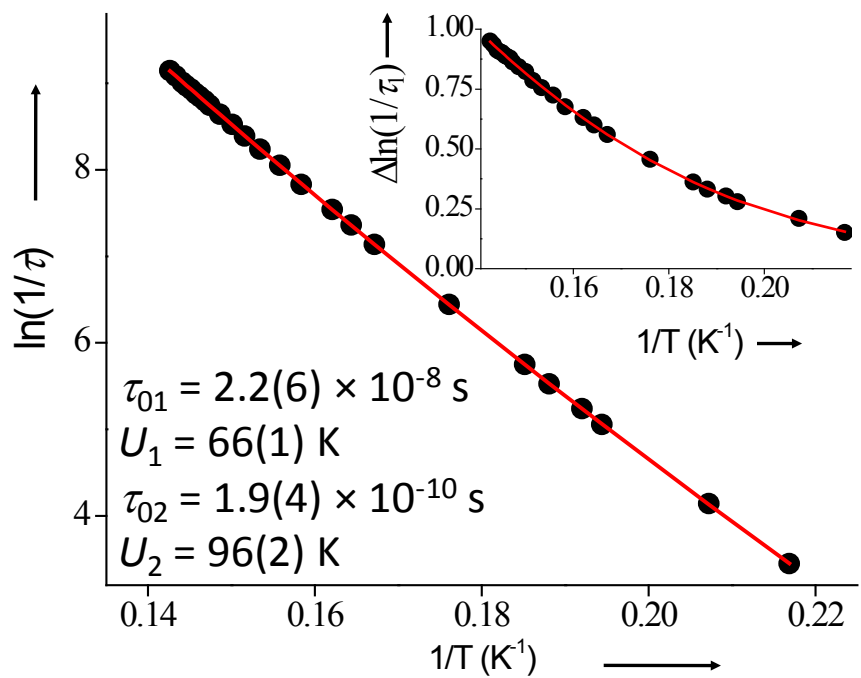

**Figure S3**. Plot of ln(1/$\tau$) vs 1/$T$ for **3**. The solid line is the fit to the double-exponential function of eq 2. The inset shows the deviation Δln(1/$\tau$) of the data from the straight line given by the first term of eq 2.

$\tau_{01}$ = 2.2(6) × 10$^{-8}$ s
$U_1$ = 66(1) K
$\tau_{02}$ = 1.9(4) × 10$^{-10}$ s
$U_2$ = 96(2) K

# COMMUNICATIONS

The magnetization relaxation of $Mn_{12}$ single-molecule magnets (SMMs) does not follow a simple Arrhenius law, even for data collected at low frequencies (5-1500 Hz) due to contributions from relaxation pathways via excited states with larger thermal barriers, but much faster attempt rates. A modified Arrhenius equation is presented as a means to overcome this problem and obtain a more reliable limiting value for the barrier to magnetization relaxation associated with the ground S=10 state.

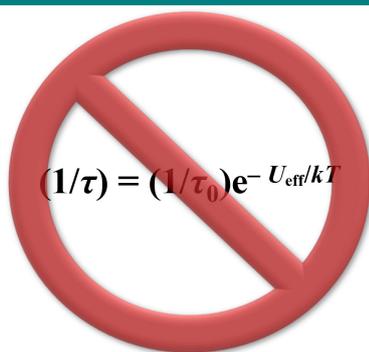

$$(1/\tau) = (1/\tau_0)e^{-U_{eff}/kT}$$

*C. Lampropoulos, S. O. Hill\*, G. Christou\**

**A Caveat for Single-Molecule Magnetism: Nonlinear Arrhenius Plots**